\documentstyle[preprint,aps]{revtex}
\begin{document}
\title{
	The Raman coupling function in amorphous silica and \\
	the nature of the long wavelength excitations in disordered
        systems.
      }
\author{
	A.~Fontana$^{1}$ ,
	R.~Dell'Anna$^{1}$ ,
	M.~Montagna$^{1}$ ,
	F.~Rossi$^{1}$ ,
	G.~Viliani$^{1}$ ,\\
	G.~Ruocco$^{2}$ ,
	M.~Sampoli$^{3}$ ,
	U.~Buchenau$^{4}$ , and
	A.~Wischnewski$^{4}$ .
       }
\address{
	 $^{1}$
	 Universit\'a di Trento and Istituto Nazionale di Fisica
	 della Materia, I-38050, Povo, Trento, Italy.\\
	 $^{2}$
         Universit\'a de L'Aquila and Istituto Nazionale di Fisica
	 della Materia, I-67100, L'Aquila, Italy. \\
	 $^{3}$
	 Universit\'a di Firenze and Istituto Nazionale di Fisica
	 della Materia, I-50139, Firenze, Italy. \\
	 $^{4}$
         Institut fur Festk\"orperforschung, Forschungszentrum, J\"ulich,
         Postfach 1913, D-52425, J\"ulich, Germany.
	}
\maketitle
\begin{abstract}
New Raman and incoherent neutron scattering data at various
temperatures and molecular dynamic simulations in amorphous silica, are
compared to obtain the Raman coupling coefficient $C(\omega)$
and, in particular, its low frequency limit. This study indicates 
that in the $\omega \rightarrow 0$
limit $C(\omega)$ extrapolates to a non vanishing value, giving 
important indications on the characteristics of the vibrational 
modes in disordered materials; in particular our results indicate that
even in the limit of very long wavelength the local disorder implies
non-regular local atomic displacements.
\end{abstract}

\newpage
\vskip .7cm

The vibrational dynamics in topologically 
disordered systems is one of the most intriguing problems
of present-day condensed matter physics \cite{and97}.
In this respect, an important issue is the
excess of vibrational states, that has been shown to
exist in glasses at low frequencies \cite{P}
by: (i) the specific heat that, above a few degrees Kelvin, 
exhibits a bump in $C_P/T^3$ \cite{SKQD,BFBCT,CDTFLSB}
and (ii) the total density of states $g(\omega)$
that, at corresponding frequencies, shows a broad excess band
in the plot of $g(\omega)/\omega^2$  \cite{BPNDAP,BZNGP,SBSFW}
often referred to as boson peak (BP). Both findings indicate a 
non-Debye behaviour of the vibrational properties at frequencies 
below 10$\div$50 cm$^{-1}$, depending
on the specific glass. The nature of the excess modes is
still object of speculations \cite{and97}. In particular, in the
case of $SiO_2$ there are two prevailing hypotheses.
According to the first one, strong phonon scattering by the structural
disorder induces a localization of the vibrational states, and the 
excess of states appears at the crossover between the low frequency 
propagating and the high frequency localized modes \cite{courtens}.
In the other picture, collective propagating modes persist up to
frequencies higher than the BP frequencies, and the BP itself reflects
their density of states \cite{sil1,sil2,sil3}.

Spectral information on the BP is usually obtained by Inelastic Neutron
Scattering (INS) or by Raman scattering (RS) spectra.
In both cases, the first order scattering intensity,  $I_{(R,N)}(\omega,T)$,
is connected to the vibrational density of states by \cite{GS}:
\begin{equation}
\label{eq1}
I_{(R,N)}(\omega,T) \propto \frac{n(\omega)+1}{\omega} g(\omega) 
C_{(R,N)}(\omega);        
\end{equation}
here $n(\omega)$ is the Bose population factor and
$C_{(R,N)}(\omega)$ is the probe-excitations coupling function.
The incoherent neutron scattering depends only on the {\it absolute} 
motion of atoms in space, so $C_{N}(\omega)=1$. On the contrary, RS 
detects the  relative displacement of neighbouring atoms and 
$C_{R}(\omega)$ turns out to be a complicated function of 
$\omega$ \cite{GS}. Since $C_R(\omega)$ is usually unknown, 
it results from Eq.~1 that RS spectra are not sufficient  
to determine $g(\omega)$, though this experimental technique 
is excellent for luminosity and resolving power. 

On a theoretical ground, different models have been developed 
for the frequency dependence of $C(\omega)$. In particular, 
it was shown that {\it i)} for slightly distorted plane wave 
vibrations, $C(\omega)\propto \omega^2$ \cite{w2}; 
{\it ii)} within the framework of the soft potential 
model \cite{GPPS}, $C(\omega)= const$; and {\it iii)} the 
fracton-like model implies $C(\omega) \propto \omega^\alpha$
\cite{FRF,FRFRD}. It is worth to note that all the models 
proposed so far, predict a power-law behaviour for $C(\omega)$.
On the contrary, the results of numerical simulations are much 
more complex and rarely produce scaling behaviour, even
on simple model systems \cite{prl90,prb91,mome,physica}.
 
On the experimental side, the direct comparison between 
RS and INS spectra appears to be a reliable procedure to 
determine $C(\omega)$, and then to check the validity 
of the proposed theoretical models. Though some attempts 
in this direction were made in the past \cite{eugPMMA}, a definite 
conclusion as to the shape of $C(\omega)$ has not been
reached \cite{SBSFW,FRFRD,ABD}, mainly due to the 
presence of "spurious" effects in RS (namely, the 
presence of quasi-elastic scattering (QES) or of 
luminescence). 

The aim of the present work is to determine the 
spectral shape of $C(\omega)$ in vitreous silica 
in the acoustic frequency range ($ \omega <150$ cm$^{-1}$), 
at different temperatures from new RS and INS measurements. 
The luminescence is carefully analysed and eliminated from 
RS spectra. The $I_R/I_N$ ratio results to be T-independent, 
except in the low frequency region where the effects of QES 
in RS are also evident. The coupling function, as defined in 
Eq.\ref{eq1} for harmonic excitations, is then extrapolated 
as the $T \rightarrow 0$ limit of $I_R/I_N$. The obtained 
function {\it does not} show the power-law behaviour previously 
proposed in the theoretical models. Further, it appears not
to vanish in the $\omega \rightarrow 0 $ limit, giving
important information on the characteristics of the normal modes
of vibration in disordered systems. These findings 
are also compared with the outcoming of a molecular dynamics 
study of vitreous silica, which confirms the experimental result
and allows a direct inspection of the vibrational modes.

The $SiO_2$ suprasil sample purchased from Goodfellow was 
used for light scattering, while Heralux quartz from Heraeus, 
was used for the neutron scattering experiments.
The Raman scattering measurements were performed using a 
standard Raman laser system. Depolarized spectra in 90$^o$ 
scattering geometry were collected in the -300$\div$5000 
cm$^{-1}$ frequency range. Such a wide range was necessary 
to take properly into account the shape and the temperature 
dependence of the luminescence background which is weak
at room temperature, but
becomes important for a quantitative determination of the 
coupling function at low frequency ($<20$ cm$^{-1}$) and 
low temperature ($<70$ K). In our case the maximum luminescence
contribution to the intensity in the problematic range
$<7$ cm$^{-1}$ was estimated
to be $\approx$ 7\% at 12 K and $\approx$ 1\% at 50 K.
The details of the data analysis
and of the luminescence subtraction will be described in a 
more extended paper \cite{tbp}. 
The neutron scattering data were taken at two different 
spectrometers (the time of flight spectrometer IN6 of 
the ILL at Grenoble, and a triple  axis spectrometer 
at the Oak Ridge High Flux Reactor). The data underwent 
the usual corrections (subtraction of the empty container, 
normalization of different detectors to a vanadium 
measurement), and the density of vibrational states
was derived following the procedure described in detail
in Ref. \cite{andbuc}.
Standard microcanonical molecular 
dynamics (MD) simulations of vitreous silica were performed 
on systems of $N=1536$ and $N=5148$ atoms (box lengths 
$L\approx$2.9 and 4.3 nm respectively) 
interacting via the two- and three-body potential proposed 
by Vashista {\it et al.} \cite{v}. The long range interaction 
was treated by the tapered reaction field method \cite{rf} 
and the equations of motion  were integrated by the leap-frog 
\cite{lf} alghoritm with a time step $\Delta t=0.5$ fs (more 
details on the MD can be found in \cite{prl98}).
We used the normal mode analysis in the harmonic
approximation and MD to derive the dynamical quantities of
interest, {\it i.e.} the density of vibrational states
$g(\omega)$ and the Raman spectrum $I_R(\omega)$. 
The computation of the Raman scattering needs a model for the 
effective atomic polarizability and its dependence on the
relative atomic position. In this study we associated a point
polarizability to each atom ($\alpha_{O}/\alpha_{Si}=4$) and
used the electric dipole propagator ($T^{(2)}(\vec r)$) to 
describe the effective atomic polarizability, 
$\Pi_i=\sum_j \alpha_i\alpha_j T^{(2)}(\vec r_i-\vec r_j)$.
Even though this polarization model may be a crude approximation
it gives a reasonable agreement with the experimental data. 
Further details of the calculation will be given in \cite{tbp}.

In Fig.~1 we report the Raman and neutron reduced spectra
$J_{(R,N)}(\omega,T)$: 
\begin{equation}
J_{(R,N)}(\omega,T)=
        \frac{I_{(R,N)}(\omega,T)}{ \omega [n(\omega)+1]},
\end{equation} 
that, according to Eq. (1), correspond to $g(\omega)C(\omega)/\omega^2$.
The data reported in Fig.~1  show that the frequency dependence of the
spectra obtained by the two techniques is similar at all the investigated
temperatures. Indeed, at room temperature, 
the maximum of the  boson peak ($\omega_{_{BP}}$) is centered at  
$\omega_{_{BP}} \approx 45$ cm$^{-1}$ and 35 cm$^{-1}$  in the RS and 
INS spectra respectively and the small blue shift of the RS 
maximum is due to the coupling function. Moreover,
the temperature dependence of $\omega_{_{BP}}$ was recently 
investigated in connection with the results of inelastic x-ray 
scattering study \cite{sil3}. By increasing $T$, the BP 
frequencies, as seen by RS and INS, increase at the same rate, 
and at the same rate as the collective excitations at $Q=1.6$ 
nm$^{-1}$ observed by inelastic X-ray scattering.
This temperature dependence was
temptatively ascribed to a next-nearest neighbour fourth-order 
anharmonicity \cite{sil3}.

The ratios of Raman to neutron data collected at different
temperatures in the range 50$\div$1100 K are reported in Fig.~2
where it is evident that the data at all temperatures 
are practically coincident for $\omega > \approx 30$ cm$^{-1}$,
while the presence of non negligible QES produces a marked $T$
dependence in the low frequency region. The QES has 
been from time to time assigned to relaxation process, 
anharmonicity or multi-phonons scattering; its origin 
is beyond the purpose of the present work.
As a consequence of QES, it is only in the
$T\rightarrow 0$ limit that the $I_R/I_N$ ratio
can be identified, apart from a proportionality factor, 
with the coupling function $C(\omega)$. However, by reducing
the temperature, also in this low frequency region the data appear 
to pile up on the continuation of the straight line that 
fits the frequency behaviour in the 30-80 cm$^{-1}$ range. The inset
in Fig. 1 shows that above $\approx$ 20 cm$^{-1}$ the 50 and 12 $K$
spectra coincide, indicating that QES is negligible at these
frequencies.

In Fig.~3, we compare the MD simulated coupling function with 
the experimental one at the lowest temperature. The MD data have 
the advantage of being free from QES contributions, but
below 30 cm$^{-1}$, owing to the finite size of the simulation box, 
the statistics is too poor to give reliable results \cite{mon}.
In the reported
range the agreement between simulated and experimental results is 
very satisfactory. As mentioned, the experimental data at $T=51$ K were 
fitted by a straight line in the 30-80 cm$^{-1}$ range (full line 
in Fig. 3). We note that our coupling function does not follow a 
simple power-law, and seems to extrapolate to a non-vanishing value 
at $\omega\rightarrow 0$ as emphasized by the dashed lines in Fig. 3 
that delimit the $\pm 2 \sigma$ confidence band of the fit.

The non vanishing value of $C(\omega  \rightarrow 0)$
gives information on the characteristics of the vibrational 
modes in topologically disordered systems.
It is well know that in crystalline materials the 
eigenvectors of the modes are plane waves, i. e. 
${\vec e}_j(p) \propto \exp(i{\vec k}_p\cdot{\vec r}_j)$ where $j$
labels the atoms and $p$ the different modes which have well 
defined wavenumbers $k_p$.
In the acoustic
region the coupling function behaves as $C(\omega) \sim \omega^2$
\cite{w2}. This $\omega^2$ law can be understood by considering 
that the scattering intensity is proportional to the square of 
the {\it relative} displacements of atoms, the main contribution 
coming from nearest neighbours; for the $p$-th mode
this relative displacement is proportional to 
$\vert {\vec e}_j(p)-{\vec e}_{j-1}(p) \vert \approx 
\vert \nabla{ e}_j(p) \vert \propto k_p$, and hence to 
$\omega_p / v$. At variance with the crystalline case, in a 
topologically disordered system the eigenvectors of the modes 
are no longer {\it pure} plane waves.
As suggested by recent MD calculations on a model LJ glass 
\cite{mazza,andv} and on a model vitreous silica \cite{TE}, 
the eigenvector of the $p$-th mode consists 
of a plane wave  part, with a rather well defined wavenumber 
$k_p$, and of a further component $\epsilon_j$, whose characteristics 
are not correlated with $k_p$, and strongly 
resembling a white "noise" in the atomic displacements,
${\vec e}_j(p)= {\vec e} \sqrt{1-\sigma^2_p} \exp(i{\vec k}_p
\cdot{\vec r}_j) + \sigma_p {\vec \epsilon}_j$. 
The plane-wave part accounts for the (rather) well
defined peak observed in the dynamical structure factor of glasses,
while the random-like part is the one mainly responsible for the value
of $C(\omega)$. Indeed, at low frequencies, the local strain associated 
with the random-like  component is much larger than that associated with the
plane wave component.
The same scenario is found in our MD simulation of v-$SiO_2$.
In Fig.~4 we report as an example the $z$ component of the eigenvector 
of a normal mode ($\omega_p \approx 80 $ cm$^{-1}$) as a function of the 
$z$ atomic coordinate in the $N=1536$ atoms system. 
The plane wave component of this mode 
corresponds to a longitudinal mode, and superimposed to it
there is the noisy part. Similar results concerning 
the characteristics of the normal modes in v-$SiO_2$ were recently
obtained by Taraskin and Elliott (Fig.~4 of ref.~\cite{TE}).
Whether the weight of this noisy component ($\sigma_p$) remains finite 
even at macroscopic wavelength, or, on the contrary, becomes negligible
at $k_p \rightarrow 0$, is still matter of discussion.
The MD, due to the finite size of the simulation box, cannot
access the very low frequeny region, and therefore no information
can be obtained on the magnitude of the noisy-to-plane wave ratio at
long wavelengths.
Indirect information on this issue can be gathered
from the behaviour of $C(\omega)$: in fact, a finite value for
$C(\omega\rightarrow 0)$, as obtained in this work, 
supports the conjecture that the noisy component {\it does not} 
disappear even in the long wavelength limit.
In fact, this can be
the only origin of a non vanishing coupling coefficient at low
frequency, because as mentioned
above, the contribution of the plane wave component to $C(\omega)$,
which is proportional to $\omega^2$, vanishes in the $\omega \rightarrow
0$ limit. 

Such conjecture is also supported by recent Brillouin light 
scattering studies of the tails of the inelatic peaks in the 
dynamic structure factors $S(Q,\omega)$ of glasses. 
In Brillouin light scattering ($Q \approx 0.02-0.04$ nm$^{-1}$)
\cite{exp1,exp2} the
$S(Q,\omega)$ shows a $\omega$-independent tail on the low frequency 
side of the inelastic peaks, whose value is $Q$-independent.
This implies that, no matter how low the $Q$ value is, the 
eigenvectors of all the modes at frequency smaller than the 
peak frequency, $\Omega_p = v Q$, contain in their space Fourier 
transform such a $Q$ component. In other words, the eigenvector of 
a mode at frequency $\Omega$, contains in its space Fourier transform
all the $Q$ components larger than $Q_p=\Omega / v$.
This is verified down to the lowest experimentally investigated
frequency ($\omega \approx 0.001$ cm$^{-1}$ \cite{exp1}) 
corresponding to a wavelength of $\approx 10$ $\mu$m.

This result seems to contradict a well established, and textbook-
reported, idea: "At wavelengths $\lambda$, much higher than the
interatomic distance $a$ and the local order correlation length $\xi$, 
the sound wave feels the disordered system as homogeneous and
all happens as in a continuous elastic medium".
The present results indicate that this picture needs to be revised
at a microscopic level.
Even when $\lambda / \xi$ is large, the modes are not pure plane waves:
the local disorder still implies non-regular local atomic displacements.

We thank S. Elliott for a preprint of Ref. \cite{TE} 
and for useful discussions.

\newpage
{\footnotesize{
\begin{center}
{\bf CAPTIONS}
\end{center}

\begin{description}

\item  {Fig. 1 - 
(a): Reduced Raman spectra, $I(\omega) / \{ \omega [ n(\omega)+1 ] \}$,
in the 0 - 150 cm${-1}$ range at different temperatures (from top to 
bottom 45, 323, 423, 523, 873, and 1073 K).
(b): Density of states divided by $\omega^2$ as obtained by inelastic
neutron scattering in the frequency range as in a) at different 
temperatures (from top to bottom  51, 318, 523, 873, and 1104 K).
The dashed line is the calculated Debye contribution 
as obtained using the sound velocities determined by 
Brillouin scattering at 50 K. (c): Reduced Raman spectra (in arbitrary
units) at 12 and 50 K
}

\item  {Fig. 2 - 
$C(\omega)$ obtained by the Raman to
Neutron scattering ratio at some significant temperatures
(from top to bottom 51 ($\circ$), 224 ($\bullet$), 271 ($\times$), 
318 ($\diamond$), 423 ($\triangle$), 523 ($\star$), 873 ($\nabla$), 
and 1104 ($+$) K).
The dashed line is the linear fit to the 51 K data in the 30-80 cm$^{-1}$ 
range. The inset shows the low frequency part of $C(\omega)$. 
}

\item  {Fig. 3 - 
Comparison between $C(\omega)$ obtained by simulation (full circles)
and experiments at $T$=51 K (open circles). The full line is the
linear fit to the experimental data in the 30-80 cm$^{-1}$ range,
and the dashed lines delimit the $\pm 2 \sigma$ confidence band.
}

\item  {Fig. 4 - 
The z-components of the eigenvector of the mode at $\omega=$ 80 cm$^{-1}$,
obtained by MD of $SiO_2$, are reported versus the z-component of the
atomic coordinates. The continuous line is the sine best fit.
}
\end{description}
}}


\begin{references}

\bibitem{and97} 
For a review see, Philosophical Magazine {\bf B77}, No. 2, (1998),
special issue: Sixth
International Workshop on Disordered Systems, Andalo, 1997, guest editors 
A.~Fontana and G.~Viliani.

\bibitem{P} 
For a review see, {\it Amorphous Solids : Low-Temperature properties}, 
edited by W.A.~Phillips, (Springer, Berlin, 1981).

\bibitem{SKQD} 
A.~P.~Sokolov, A.~Kisliuk, D.~Quitmann, and E.~Duval, 
Phys. Rev {\bf B48}, 7692 (1993).

\bibitem{BFBCT} 
A.~Brodin, A.~Fontana, L.~Borjesson, G.~Carini and L.M.~Torell, 
Phys. Rev. Lett. {\bf 73}, 2067 (1994).

\bibitem{CDTFLSB} 
G.~Carini {\it et al.}, 
Phys. Rev. {\bf B52}, 9342 (1995). 

\bibitem{BPNDAP} 
U.~Buchenau {\it et al.}, 
Phys. Rev. {\bf 34}, 5665 (1986).

\bibitem{BZNGP} 
U.~Buchenau, H.M.~Zhou, N.~Nucker, K.S.~Gilroy, and W.A.~Phillips,
Phys. Rev. Lett. {\bf 60}, 1318 (1988).

\bibitem{SBSFW} 
A.P.~Sokolov, U.~Buchenau, W.~Steffen, B.~Frick, and A.~Wischmewsski,
Phys. Rev. {\bf B52}, R9815 (1995).

\bibitem{courtens} 
M.~Foret, E.~Courtens, R.~Vacher, J.B.Suck,
Phys. Rev. Lett. {\bf 78}, 4670 (1997).

\bibitem{sil1} 
P.~Benassi {\it et al.},  
Phys. Rev. Lett., {\bf 77}, 3835 (1996).

\bibitem{sil2} 
C.~Masciovecchio {\it et al.}, 
Phys. Rev. {\bf B55}, 8049 (1997).

\bibitem{sil3} 
C.~Masciovecchio {\it et al.}, 
preprint.

\bibitem{GS} 
F.L.~Galenneer, and P.L.~Sen, 
Phys. Rev. {\bf B17}, 1928 (1978).

\bibitem{w2} 
A.J.~Martin and W.~Brenig, 
Phys. Status Solidi (B) {\bf 64}, 163 (1972).

\bibitem{GPPS} 
V.L.~Gurevich, D.A.~Parshin, J.~Pelous, and H.R.~Schober, 
Phys. Rev. {\bf B48}, 16318 (1993).

\bibitem{FRF} 
A.~Fontana, F.~Rocca, and M.P.~Fontana, 
Phys. Rev. Lett. {\bf 58}, 503 (1987).

\bibitem{FRFRD} 
A.~Fontana, F.~Rocca, M.P.~Fontana, B.~Rosi, and A.J.~Dianoux, 
Phys. Rev. {\bf B41}, 3778 (1990).

\bibitem{prl90}
M.~Montagna {\it et al.}, 
Phys. Rev. Lett. {\bf 65}, 1136 (1990).

\bibitem{prb91}
V.~Mazzacurati {\it et al.}, 
Phys. Rev. {\bf B45}, 2126 (1992).

\bibitem{mome}
G.~Viliani {\it et al.}, 
Phys. Rev. {\bf B52}, 3346 (1995).

\bibitem{physica}
O.~Pilla, G.~Viliani, R.~Dell'Anna, and G.~Ruocco, Physica {\bf A247},
23 (1997).

\bibitem{eugPMMA}
A.~Mermet, N.V.~Surovtsev, E.~Duval, J.F.~Jal, and A.J.~Dianoux,
Europhys. Lett. {\bf 36}, 277 (1996).

\bibitem{ABD} 
T.~Achibat, A.~Boukenter, and E.~Duval, 
J. Chem. Phys. {\bf 99}, 2046 (1993).

\bibitem{tbp} 
M.~Montagna {\it et al.}, to be published.

\bibitem{andbuc} 
A.~Wischnewski, U.~Buchenau, A.J.~Dianoux, W.A.~Kamitakahara,
and J.L.~Zaretsky, Phil. Mag. {\bf B77}, 579 (1998).

\bibitem{v} 
P.~Vashista, R.K.~Kalia,  J.P.~RinoD, and I.~Ebbsj\"o, 
Phys. Rev.  {\bf B41}, 12197 (1990).


\bibitem{rf} 
G.~Ruocco and  M.~Sampoli, Molec. Phys. {\bf 82}, 875 (1994).

\bibitem{lf}
M.P.~Allen and D.J.~Tildesley, 
{\it Computer Simulation of Liquids},
Oxford University Press,  (N.Y. 1990).

\bibitem{prl98} 
R.~Dell'Anna, G.~Ruocco, M.~Sampoli, and G.~Viliani, 
Phys. Rev. Lett {\bf80 }, 1236 (1998).

\bibitem{mon}
M.~Montagna, P.~Benassi, W.~Frizzera, O.~Pilla, G.~Viliani,
V.~Mazzacurati, G.~Ruocco, G.~Signorelli,
Physica {\bf A191}, 348 (1992)



\bibitem{mazza} 
V.~Mazzacurati, G.~Ruocco and M.~Sampoli,
Europhys. Lett. {\bf 34}, 681 (1996).

\bibitem{andv} 
M.~Sampoli, P.~Benassi, R.~Dell'Anna, V.~Mazzacurati, and G.~Ruocco, 
Phil. Mag. {\bf B77}, 473 (1998).

\bibitem{TE}                                         
S.N.~Taraskin, and S.R.~Elliot, preprint.

\bibitem{exp1} 
G.~Monaco, G.~Ruocco, L.~Comez, and D.~Fioretto, 
J. Non Cryst. Sol. {\bf 00}, 0000 (1998);
G.~Monaco, D.~Fioretto, C.~Masciovecchio, G.~Ruocco, and F.~Sette, 
preprint.

\bibitem{exp2} 
P.~Benassi, A.~Fontana, V.~Mazzacurati, and M.~Sampoli, 
preprint.

\end{references}
\end{document}